%% file: morphogenesis13_clean_arxiv.tex
\documentclass[aps,pre,twocolumn,superscriptaddress,longbibliography]{revtex4-1}

\usepackage{amsmath}
\usepackage{graphicx}
\usepackage{mwe}
\usepackage[caption=false]{subfig}
\usepackage[usenames, dvipsnames]{color}
\usepackage{verbatim}
\usepackage{soul,xcolor}
\usepackage{enumitem}

\begin{document}
%\setstcolor{red}

% Use the \preprint command to place your local institutional report
% number in the upper righthand corner of the title page in preprint mode.
% Multiple \preprint commands are allowed.
% Use the 'preprintnumbers' class option to override journal defaults
% to display numbers if necessary
%\preprint{}

%Title of paper
\title{Buckling without bending: a new paradigm in morphogenesis}

% repeat the \author .. \affiliation  etc. as needed
% \email, \thanks, \homepage, \altaffiliation all apply to the current
% author. Explanatory text should go in the []'s, actual e-mail
% address or url should go in the {}'s for \email and \homepage.
% Please use the appropriate macro foreach each type of information

% \affiliation command applies to all authors since the last
% \affiliation command. The \affiliation command should follow the
% other information
% \affiliation can be followed by \email, \homepage, \thanks as well.
%
%
\author{T. A. Engstrom}
\email[]{tyler.engstrom@gmail.com}
\affiliation{Department of Physics, Syracuse University, Syracuse, NY 13244, USA}
\author{Teng Zhang}
\email[]{tzhang48@syr.edu}
\affiliation{Department of Mechanical \& Aerospace Engineering, Syracuse University, Syracuse, NY 13244, USA}
\author{A. K. Lawton}
\affiliation{Memorial Sloan Kettering Cancer Center, New York, NY 10065, USA}
\author{A. L. Joyner}
\affiliation{Memorial Sloan Kettering Cancer Center, New York, NY 10065, USA}
\author{J. M. Schwarz}
\email[]{jschwarz@physics.syr.edu}
\affiliation{Department of Physics, Syracuse University, Syracuse, NY 13244, USA}

%Collaboration name if desired (requires use of superscriptaddress
%option in \documentclass). \noaffiliation is required (may also be
%used with the \author command).
%\collaboration can be followed by \email, \homepage, \thanks as well.
%\collaboration{}
%\noaffiliation

\date{\today}

\begin{abstract}
A curious feature of organ and organoid morphogenesis is that in certain cases, spatial oscillations in the thickness of the growing ``film" are out-of-phase with the deformation of the slower-growing ``substrate," while in other cases, the oscillations are in-phase. The former cannot be explained by elastic bilayer instability, and contradict the notion that there is a universal mechanism by which brains, intestines, teeth, and other organs develop surface wrinkles and folds. Inspired by the microstructure of the embryonic cerebellum, we develop a new model of 2d morphogenesis in which system-spanning elastic fibers endow the organ with a preferred radius, while a separate fiber network resides in the otherwise fluid-like film  at the outer edge of the organ and resists thickness gradients thereof. The tendency of the film to uniformly thicken or thin is described via a ``growth potential". Several features of cerebellum, +blebbistatin organoid, and retinal fovea morphogenesis, including out-of-phase behavior and a film thickness amplitude that is comparable to the radius amplitude, are readily explained by our simple analytical model, as may be an observed scale-invariance in the number of folds in the cerebellum. We also study a nonlinear variant of the model, propose further biological and bio-inspired applications, and address how our model is and is not unique to the developing nervous system.

\end{abstract}

% insert suggested PACS numbers in braces on next line
%\pacs{46.25.-y, 46.35.+z, 62.20.mq, 68.35.Rh, 82.60.Nh, 87.15.Zg} 
% insert suggested keywords - APS authors don't need to do this
%\keywords{}

%\maketitle must follow title, authors, abstract, \pacs, and \keywords
\maketitle

% body of paper here - Use proper section commands
% References should be done using the \cite, \ref, and \label commands

%%%
\section{Introduction}
%%%

An elastic instability, driven by differential growth, is thought to be broadly responsible for many of the motifs seen in organ morphogenesis~\cite{nel16}. Indeed, this mechanism has been studied in the context of brain folds~\cite{ric75, rag97, bay13, man14, tal14, bud15, tal16, lej16, kar18}, intestinal crypts and villi~\cite{han11, shy13}, airway mucus wrinkles~\cite{wig97, li11}, tooth ridges~\cite{osb08}, and hair follicle patterns~\cite{shy17, sil18}, among others. Wrinkling or buckling provides a means for these organs' shapes to emerge reliably from their respective starting geometries, without appealing to spatial variation in gene expression or other biochemical pre-patterning of folds. Apart from geometry, all that is required is a competition between the bending energy of a uniformly growing film (cortex/mucus/epithelium) and the energy to stretch and compress a slower growing substrate (subcortex/submucus/mesenchyme). 

Biological tests of wrinkling predictions have largely focused on pattern formation and wavelength, the latter scaling with film thickness and power 1/3, 1/4, or 1/6 of the stiffness contrast, depending on the substrate model~\cite{cer03}. (Stiffness contrast is defined as Young's modulus of the film divided by that of the substrate.) The wavelength test is typically not a conclusive test for two reasons. First, measuring the stiffness contrast \emph{in vivo}, at the appropriate developmental period, and on the timescale relevant to morphogenesis, is technically challenging. To our knowledge it has not been done, and the best available data come from \emph{ex vivo} measurements on freshly harvested embryonic organs~\cite{shy13, wei17}. Second, the weak dependence of wavelength on stiffness contrast exacerbates the uncertainty of the latter: a large range of stiffness contrasts might be said to ``agree" with an observed wavelength.  Considering these difficulties, it is surprising that other wave properties predicted by wrinkling theory have not been given more attention. In particular, the phase and amplitude behaviors associated with elastic wrinkling are qualitatively distinct and constitute a stringent pass/fail test with regard to experimental biology observations. Let us see how this works.

\begin{figure}[t]
\centering
\includegraphics[width=0.48\textwidth]{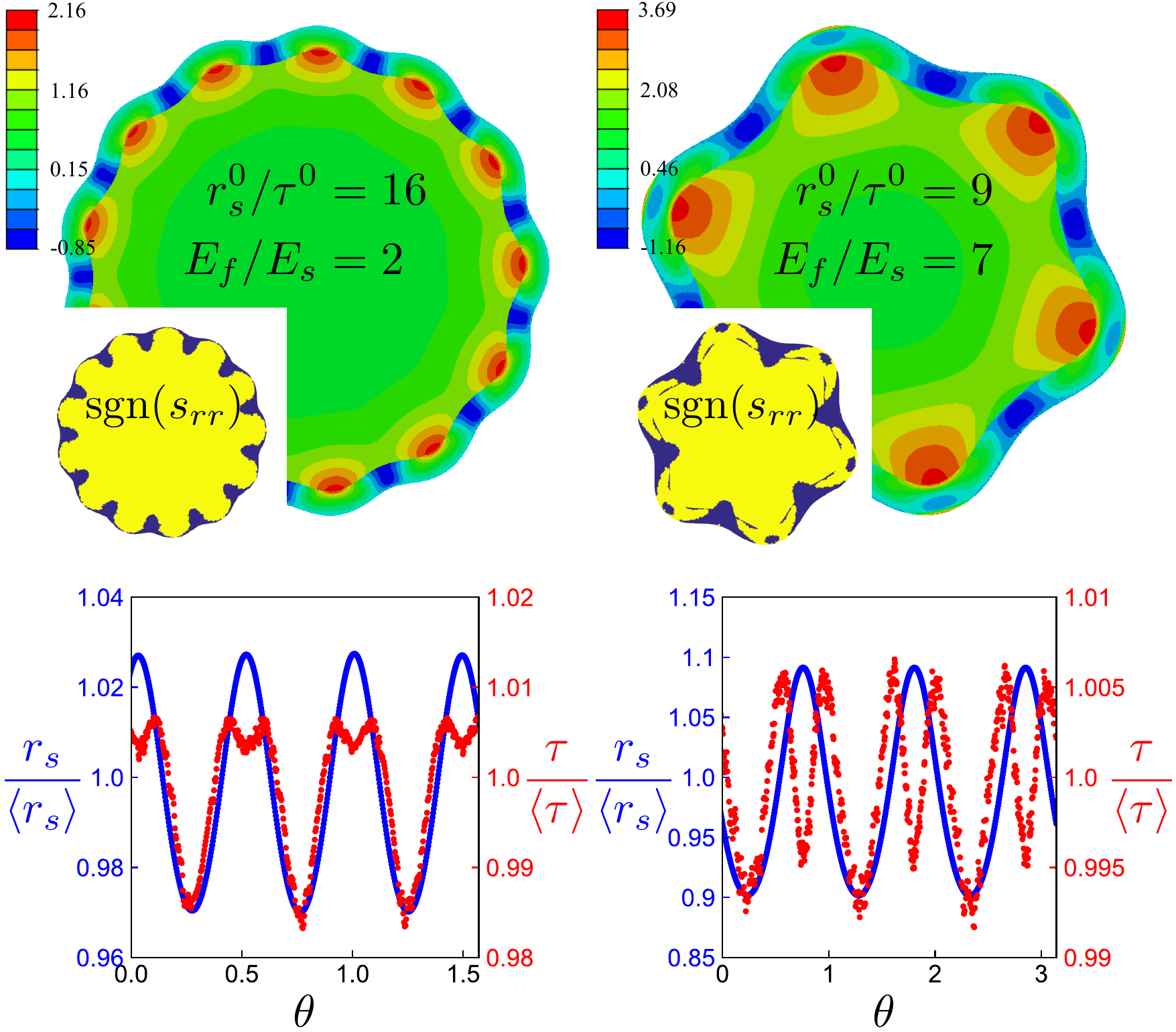}
\caption{\label{circular_wrinkle}Elastic bilayer wrinkling generates film thickness oscillations that are overall in-phase with the substrate deformation when the film is thin. Top row: maximum in-plane principal stress (normalized to half of the substrate's shear modulus) for two different circular wrinkling problems. $r_s^0/\tau^0$ is the initial substrate radius in units of the initial film thickness, and $E_f/E_s$ is the stiffness contrast. Insets plot the sign of the stress tensor component $s_{rr}$, with yellow indicating positive values (tension) and purple indicating negative values (compression). Bottom row: substrate radius (blue) and film thickness (red), normalized to their average values, for the same two wrinkling problems. Note the different scales on left and right axes.}
\end{figure}

In simple wrinkling analyses, e.g., Euler buckling of a film adhered to a bed of springs which yields the power 1/4 mentioned earlier, the film thickness is typically assumed to be spatially uniform. In reality, the film thickness is modulated by the substrate deformation. A crude consideration of the forces acting on the quasi-static interface indicates the film thickness oscillations should be \emph{in-phase} with the substrate deformation. That is, thick spots in the film (regions of the film under internal tension) should be matched up with thick spots in the substrate (regions of the substrate under internal tension), and thin spots with thin spots. Finite element simulations confirm the essential correctness of this argument for wrinkling in planar geometry, and also circular geometry, provided the film is thin compared to the substrate radius and the stiffness contrast is not too large (see Figure~\ref{circular_wrinkle}). In general, however, simulations reveal two film thickness minima per wrinkle: one coincides with the wrinkle valley, and the other coincides with the wrinkle crest, as can be seen in the lower panels of Figure~\ref{circular_wrinkle}. The latter has small depth and width compared to the former (i.e., the thickness profile is overall in-phase) except when the modulus ratio is large and/or the film thickness is a significant fraction of the substrate radius. Generically, the amplitude of thickness oscillations is much smaller than the wrinkling amplitude. The large thickness limit that one might expect to yield comparable amplitudes tends not to undergo wrinkling, but rather a global buckling mode~\cite{lag16}. Full details of these simulations, as well as additional examples (one of which shows a wrinkling + global buckling mixed mode), are given in the Appendix.
%
% Figure 1
%

The simple quasi-statics argument for in-phase behavior straightforwardly extends to growth and elastic modulus profiles (as a function of radius) that are more complicated than a simple bilayer profile, including even the possibility of non-monotonicity. First notice that any smooth, continuous growth and/or modulus profile can be represented by many discrete layers. Specializing to those layers having thicknesses that are small compared to their radii, we observe that any two adjacent such elastic layers must exert equal and opposite normal forces at every point along their shared interface. As before, normal tension and normal compression correspond to thicker than average, and thinner than average, respectively, assuming initial axi-symmetry. Thus, where any one layer is thicker than its average thickness, all the other layers must be relatively thick, and any arbitrary contiguous grouping must be relatively thick. This generalization serves to emphasize that in-phase thickness behavior is a generic consequence of differentially-growing elastic multilayers (assuming near-planar geometry), and is independent of the precise details of the growth and/or modulus profiles.

\begin{figure}[t]
\centering
\includegraphics[width=0.46\textwidth]{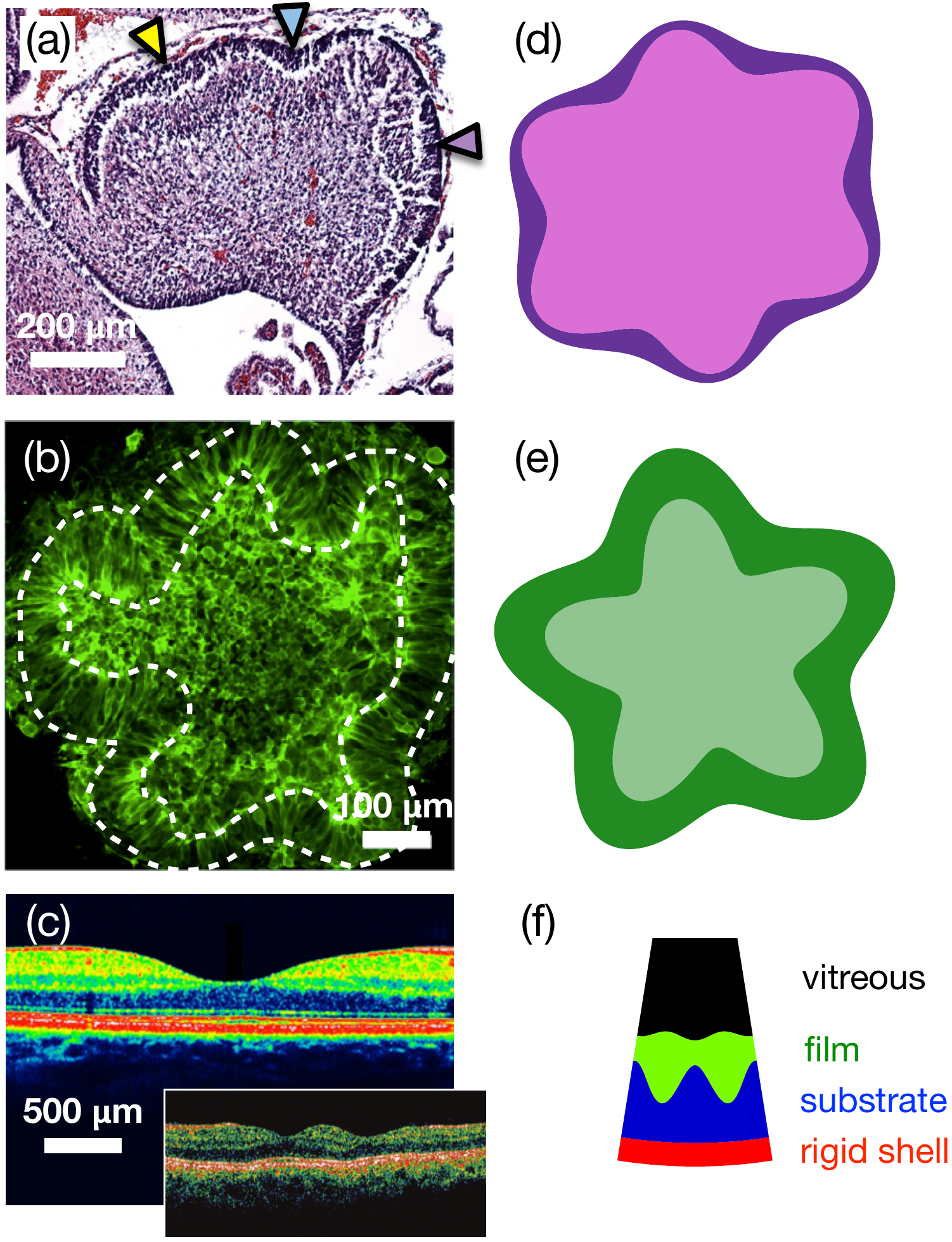}
\caption{\label{bio_systems}Certain morphogenesis problems exhibit out-of-phase thickness oscillations, at odds with an elastic wrinkling mechanism. (a) Midsaggital section of a mouse cerebellum at 17.5 embryonic days, reprinted from Fig 5 in Ref.~\cite{joy17} with permission from Springer Nature. Arrows are from the original image and mark positions of developing invaginations. (b) +Blebbistatin organoid, reprinted from Fig 4a in Ref.~\cite{kar18} with permission from Springer Nature. (c) Cross section of the foveal pit in a human retina, reprinted from Fig 13b in Ref.~\cite{kol18} with permission from Webvision and the author. The green and blue layers are the ganglion cell and photoreceptor layers, respectively. Inset shows a rare double fovea in a human retina (double foveas are typical in some bird species), reprinted from Ref.~\cite{beh07} with permission from the American Medical Association. (d) Polar plot of $r(\theta)$ and $r(\theta)-t(\theta)$ from Equations~\ref{tsol}-\ref{rsol} with $\epsilon=0.6$, $c=T/a=t_0/a=0.1$, $k_t/\beta=31.3$, $\phi=\pi$, $e=0.5$. (e) Same as (d) but with $\epsilon=0.9$, $c=0.067$, $T/a=0.05$, $t_0/a=0.7$, $k_t/\beta=15.6$, $\phi=e=0$. (f) Polar plot of $r(\theta)$ and $r(\theta)-t(\theta)$ from Equations~\ref{tsol_concave}-\ref{rsol_concave} with $\epsilon=-0.2$, $c=0.3$, $T/a=t_0/a=1$, $R/a=20$, $k_t/\beta=1770$, $\phi=e=0$.}
\end{figure}

So while elastic instabilities driven by differential growth exhibit a number of interesting features, it follows, that if a differentially growing biological system were found with quasi-static \emph{out-of-phase} behavior (film thickness maxima coinciding with substrate valleys, and thickness minima with hills), elastic wrinkling could potentially be ruled out as the mechanism of shape change. Moreover, a thickness amplitude that is not small compared to the surface height amplitude would be at odds with elastic wrinkling.  And yet it is these very ``anti-wrinkling" behaviors that show up in several motifs in organ morphogenesis: the cerebellum, certain brain-like organoids grown \emph{in vitro}, and the retinal fovea (see Figure~\ref{bio_systems}a-c). Note that all three examples are pertinent to nervous system development. The first two of these have been previously considered as elastic wrinkling problems~\cite{lej16, kar18}, as has brain folding more generally~\cite{ric75, rag97, bay13, man14, tal14, bud15, tal16}. Fovea formation has not previously been credited to elastic instability, as far as we know, but there does appear to be differential growth between the constituent layers of the retina in the vicinity of the developing fovea, e.g., Figure 2 in Ref.~\cite{hen92}. In effect, the first two of these biological systems, and tentatively the third about which less is known, are counter-examples to the elastic instability paradigm for morphogenesis. They call for a new paradigm.
%
% Figure 2
%

In constructing this new paradigm for shape change in developing organs, we go beyond modeling the constituent layers as elastic materials with different elastic constants. If one looks at microstructures within the developing nervous system, one typically finds globular-cells, such as granular cell precursors in the cerebellum or neural precursors in the cerebrum, and fiber-like cells, such as radial glial.  At some length-scale both types of cells, if they are interconnected enough, can presumably be modeled as an elastic continuum.  But at what length-scale and degree of interconnectedness is such modeling justified? The microstructure of the developing organ hands us a clue as to when such an approximation breaks down and so we should look to it in developing this new paradigm. Moreover, it has recently been shown that in a model for confluent cell tissue, the introduction of cell division drives the tissue from an elastic solid to a fluid~\cite{mat17}. As with any developing mammalian organ, cell division is tantamount to the process, so perhaps we must further relax the notion that the different components involved are all elastic and begin to consider that at least some of the components are fluid-like.

Because the new paradigm is rooted in the microstructure of a developing organ, we adopt the embryonic day 16.5-18.5 mouse cerebellum, with its fast-growing outer cortex (more precisely, the external granular layer) and slow-growing inner core, as our prototype system. One of us previously documented the out-of-phase, ``anti-wrinkling" character of this system~\cite{sud07}, and it is the subject of further investigations parallel to this work~\cite{law18}. In the two day window just mentioned, the mouse cerebellum transitions from having a featureless convex surface, to developing smooth, sinusoidal radius oscillations, to then forming cusped invaginations, termed ``anchoring centers". The anchoring centers delineate smooth outward protuberances called lobules~\cite{joy17}. Later in development, some of the first generation lobules subdivide into second generation lobules~\cite{sil07}. (Human cerebella appear to have several generations of subdivisions.) In the following, we will exploit the quasi-2d nature of the cerebellum (its rows of parallel folds indicate that most of the biomechanical action is in the parasaggital plane), which makes it a simpler system to analyze than the cerebrum with its 3d folds. We note that earlier work on modeling the cerebral cortex as a smectic liquid crystal (as opposed to a purely elastic solid), being pulled on by axonal tension, is a first step in the direction of incorporating more of the microstructure into cerebral development~\cite{man14}. 

As for additional applications of the model, a recent \emph{in vitro} experiment demonstrates shape change in a human embryonic stem cell-derived aggregate inserted into a quasi-2d microfabricated compartment filled with hydrogel~\cite{kar18}.  After several days, elongated fiber-like cells are present along the periphery of the aggregate, while globular-like cells remain in the interior. The former are representative of a cortex and the latter, an inner core. During its second week in development, the surface of the organoid invaginates in a manner reminiscent of brain folds. The introduction of blebbistatin, a myosin-inhibitor, produces qualitatively different shape changes as compared to the untreated case.  We will apply our prototype model to the +blebbistatin organoid, as opposed to the untreated case, since the developing shapes in +blebbistatin organoid more closely resemble the shapes obtained in the model. We will discuss the untreated case as a potentially more sophisticated model.  

Finally, we consider retinal development as a third possible application of (a variant of) our prototype model. The developing retina is encased in a rigid shell, with several distinct layers of cells supported by the inner part of the shell, followed by vitreous, a clear gel, filling the space between these layers and the lens of the eye~\cite{kol18}.  M\"uller glial fibers span some of the layers in the macular region, where the center of the field of vision is focused.  A depression known as the foveal pit begins to form in this region around human gestational week 25, and develops over a time-scale comparable to that of brain folding~\cite{hen92}.

%%%
\section{Model}
%%%

%
% Figure 3
%
\begin{figure}[t]
\centering
\includegraphics[width=0.4\textwidth]{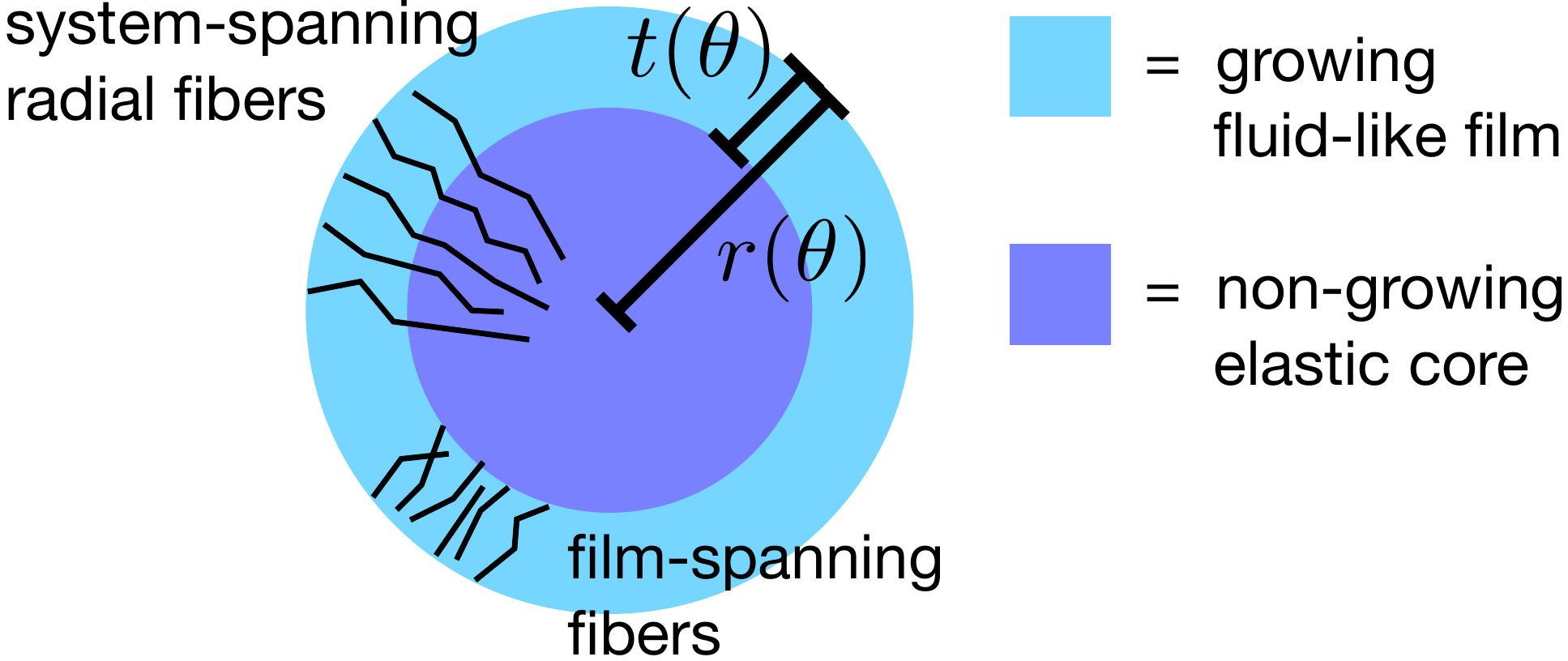}
\caption{\label{schematic}Idealized, parasaggital section of an embryonic cerebellum. The cortex is modeled as a growing fluid-like ``film". Bergmann glia fibers span this film, while radial glia fibers span the structure. Periodic boundary conditions are likely not relevant to a real cerebellum, whose cortex is discontinuous owing to a ventricular zone and attachment to the brain stem, but they will be adopted here for simplicity.}
\end{figure}
Let us model the growing cerebellar cortex as a 2d annulus-like region having outer radius $r$ and thickness $t$, which are scalar functions of an angular coordinate $\theta$. In other words, $t$ is \emph{defined} to be measured in the radial direction (see Figure~\ref{schematic}). This simple parameterization is valid only for weak deviations from an annulus. Consider, for example, that a deep or overhanging surface fold could generate multivalued $r(\theta)$, as well as lead to a $t(\theta)$ that violates one's sensibilities around the usual notion of thickness. Under this restriction, we introduce the quasi-static energy functional
\begin{equation}
E[r,t,\tfrac{dt}{d\theta}] = \int d\theta \Big{\{} k_r(r-r_0)^2 - k_t(t-t_0)^2 + \beta\Big(\frac{dt}{d\theta}\Big)^2 \Big{\}},\label{E}
\end{equation}
to be minimized subject to a constraint on the area of the non-growing subcortex, i.e.,
\begin{equation}
\frac{1}{2}\int d\theta(r-t)^2=A_0=\textrm{constant}.\label{constraint}
\end{equation}
The variational problem at hand is thus 
\begin{equation}
\delta\Big(E-\mu\int d\theta(r-t)^2\Big)=0,\label{varPrinc}
\end{equation}
where $\mu$ is a Lagrange multiplier whose value will be determined upon simultaneous solution of Equations~\ref{constraint}-\ref{varPrinc}. 

In Equation~\ref{E}, $k_r$, $k_t$, and $\beta$ are all positive constants. The first term encodes a preferred radius $r_0$, or more generally, a preferred shape $r_0(\theta)$. Due to its negative contribution, the second term favors thickening (or thinning) of the annulus with respect to a reference thickness $t_0$, and for simplicity, we will take $t_0$ to be a constant. Thus while $k_r$ is a modulus, $k_t$ can be regarded as a ``growth potential," and the corresponding terms compete with one another because of the subcortex incompressibility. This competition tends to drive the system away from its preferred shape. The third term in Equation~\ref{E} penalizes spatial variations in thickness -- it is \emph{not} a bending term as its appearance may suggest to some. (A bending term would involve a squared second derivative of the film deflection.) In fact, the absence of a cortex bending modulus is a key feature of the present model that distinguishes it from elastic bilayer models of brain folding. Here the cortex resembles a mixture of fluid + fibrous scaffolding held inside a container. Both container and scaffolding are flexible in bending, but the scaffolding resists gradients in the container's thickness. As for the physical nature of this scaffolding in an actual cerebellum, we suggest its main component is the cortex-spanning Bergmann glia, depicted schematically in Figure~\ref{schematic}. The non-growing subcortex, in contrast, is treated as an elastic solid. Together with radially-oriented fibers that span the system (radial glia), this is the origin of the preferred radius term in Equation~\ref{E}. We suggest that radial glia ``tether" the surface to the subcortex, using their washer-like pial endfeet to distribute load over the flexible surface, while their other ends remain securely anchored in the solid subcortex. Implicit in this anchoring mechanism is the subcortex's finite shear modulus, however, the actual shear deformation energy of the subcortex is neglected.

\begin{figure}[b]
\centering
\includegraphics[width=0.35\textwidth]{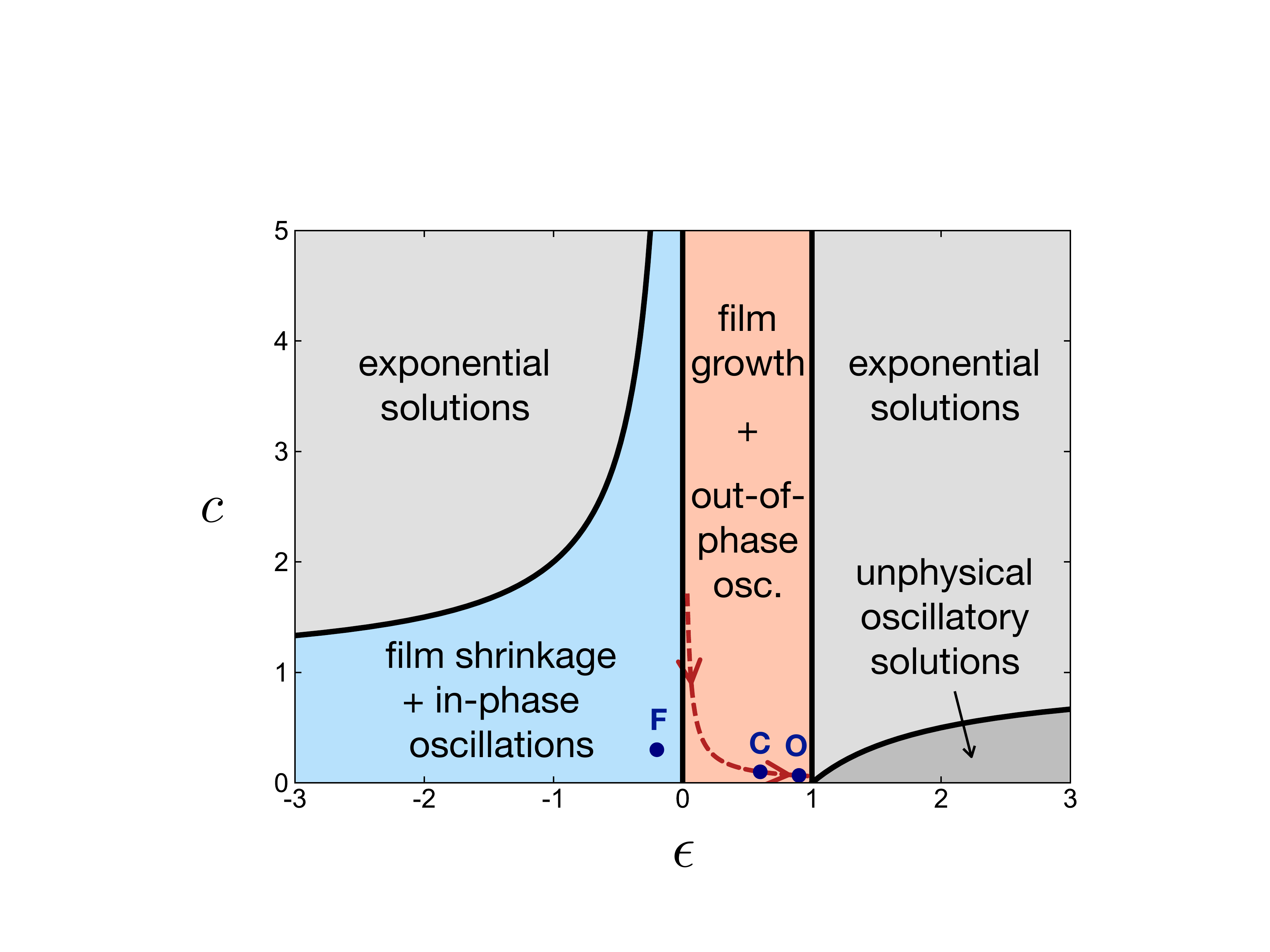}
\caption{\label{pd}Phase diagram of behavioral regimes, where $\epsilon=\mu/k_r$ and $c=k_r/k_t$. Phase boundaries are represented by solid black lines; those that are curved obey $c=1-\epsilon^{-1}$. Blue and orange shaded regions may pertain to morphogenesis. Dark blue dots labeled C, O, and F indicate parameter values used to make the cerebellum, organoid, and fovea plots, respectively, in Figure~\ref{bio_systems}. The red dashed line (given by $c=0.06/\epsilon$) indicates one possible trajectory of cerebellar and organoid morphogenensis, associated with $k_r$ decreasing over time.}
\end{figure}

So that we may make progress analytically, let us take the preferred shape to be an ellipse having semi-major axis $a$ and (small) eccentricity $e$, i.e.,
\begin{equation}
r_0(\theta)=a\big(1-\tfrac{e^2}{2}\sin^2\theta\big).
\end{equation}
With this choice, decoupling the Euler-Lagrange equations results in an unconventionally driven oscillator equation for the thickness
\begin{equation}
t''(\theta) + q^2t(\theta) = f_0 - f_1\sin^2\theta,\label{ODE}
\end{equation}
where $q^2=\frac{k_t}{\beta}(1+\frac{\epsilon c}{1-\epsilon})$, $f_0=\frac{k_t}{\beta}(t_0+\frac{\epsilon c a}{1-\epsilon})$, $f_1=\frac{k_t}{\beta}(\frac{\epsilon c a}{1-\epsilon})\frac{e^2}{2}$, $\epsilon=\mu/k_r$, and $c=k_r/k_t$ are all constants (within the quasi-static description). In the regions of the $\epsilon$-$c$ plane where $q$ is real (see Figure~\ref{pd}), Equation~\ref{ODE} has the general solution
\begin{eqnarray}
t(\theta) &=& T\sin(q\theta + \phi) + \frac{(1-\epsilon)t_0+\epsilon ca}{1-\epsilon+\epsilon c} \nonumber\\
&+& \frac{\epsilon}{1-\epsilon}\bigg(\frac{k_rae^2}{2\beta}\bigg)\bigg(\frac{2-q^2\sin^2\theta}{(2-q^2)^2-4}\bigg),\label{tsol}
\end{eqnarray}
in terms of which the radius is
\begin{equation}
r(\theta) = \bigg(\frac{1}{1-\epsilon}\bigg)r_0(\theta) - \bigg(\frac{\epsilon}{1-\epsilon}\bigg)t(\theta).\label{rsol}
\end{equation}
Combining Equations~\ref{constraint}, \ref{tsol}, and \ref{rsol}, we find the thickness amplitude is given by
\begin{equation}
T = \sqrt{2}(1-\epsilon)\sqrt{\tfrac{A_0}{\pi} - \big(\tfrac{a-t_0}{1-\epsilon+\epsilon c}\big)^2\big(1-\tfrac{a}{a-t_0}\tfrac{e^2}{2} + \mathcal{O}(e^4)\big) },\label{amplitude}
\end{equation}
and evidently we must impose a lower bound on $A_0$ to ensure this amplitude is real-valued:
\begin{equation}
A_0 > \frac{\pi(a-t_0)^2}{(1-\epsilon+\epsilon c)^2} \bigg(1 - \frac{a}{a-t_0}\frac{e^2}{2} + \mathcal{O}(e^4)\bigg).\label{lowerbound}
\end{equation}

The physical significance of the Lagrange multiplier's sign is illuminated by a relationship between angle-averaged quantities
\begin{equation}
\langle t-t_0\rangle = \epsilon c \langle r-t\rangle,\label{grow_shrink}
\end{equation}
which is exact to all orders in $e$. Since physically viable solutions require $r-t>0$ for all $\theta$, Equation~\ref{grow_shrink} says that $\epsilon>0$ corresponds to growth of the cortex while $\epsilon<0$ corresponds to shrinkage (not in a dynamical sense, necessarily, but with respect to the quasi-static value of $t_0$). Note $\langle r-t\rangle>0$ also implies that oscillatory solutions having $\epsilon>1$ are presumably unphysical in the sense $a-t_0<0$. Consequently we will restrict our focus to $\epsilon<1$. 

Apart from its sign, the Lagrange multiplier may be thought of as a pressure-like quantity, or a ``chemical potential" for changing the core area by a unit amount (the symbol $\mu$ was deliberately chosen to suggest this analogy). It can also be argued that $\epsilon$ should set the amplitude $T$. Consider that as $\epsilon\to0$, the competition between preferred radius and film growth disappears, as nothing prevents the film from growing uniformly inward and consuming the core. Absent this competition, there is no driving force for film thickness oscillations, one might argue. The simplest way to make the model consistent with this argument would be to set $T\sim\epsilon$, i.e., make the film thickness amplitude linearly go to zero as $\epsilon$ goes to zero (the radius amplitude would go to zero quadratically). This assignment would also establish $\epsilon$ as the key parameter governing all phase and amplitude aspects of shape-change. Endowing $\epsilon$ with time-dependence would therefore be a natural (and minimal) starting point for dynamical morphogenesis problems, within the current modeling framework. For the remainder of this paper, however, we shall mostly confine ourselves to a statics approach, as well as regard $T$ and $\epsilon$ as independent fitting parameters. Time-dependence of the model parameters is further investigated elsewhere~\cite{law18}.   

The simplicity of the above-described model belies its novelty, in that several of the predicted behaviors are opposite those predicted by differentially-growing, elastic bilayer models. For example, $t$ and $r-t$ oscillations are always out-of-phase when $\epsilon<1$ (compare $\tau$ and $r_s$ oscillations, respectively, in Figure~\ref{circular_wrinkle}), while $t$ and $r$ oscillations can be either in-phase, or out-of-phase, depending upon position in the $\epsilon$-$c$ plane (see Figure~\ref{pd}). Likewise, the amplitude of $t$ oscillations can be either greater than, or less than, the amplitude of $r$ oscillations (for $\epsilon<1/2$ and $\epsilon>1/2$, respectively). As another example, notice that as $\epsilon\to0$, the wavenumber $q$ depends only on a ratio of energy scales associated with microscopic mechanisms, i.e., $k_t/\beta$. In contrast, elastic wrinkling predicts that the number of waves depends on a ratio of length scales, roughly equivalent to our $a/t_0$, which may or may not be scale-invariant.
%
% Figure 4
%

In Figure~\ref{bio_systems}d-e, we compare plots of $r$ and $r-t$ with images of an embryonic mouse cerebellum and a brain-like organoid treated with blebbistatin. (The region in-between the two plotted curves, representing the cortex, is filled with dark purple and dark green, respectively.) Some comments specific to the organoid application are now in order. As previously noted, the inner core of this structure consists of globular-like cells, while the periphery consists of fibrous cells as well as motile cells that divide and move along the fibrous cells in such a way that the periphery grows faster than the core. The fibrous cells and motile cells are reminiscent of Bergmann glia and granular precursor cells in the cerebellum, respectively.  There are assumed to be radial glial cells or some other means of transmiting radial tension throughout the organoid, such as that discussed in Ref~\cite{law18}. From this description, one could argue that the model at hand may be applicable to the untreated organoid.  However, the model does not take into account the active contractility of the core, due to the myosin that is present there. This activity is likely to be mechanosensitive such that assumptions beyond those we have already made would be required. While this is an interesting avenue to pursue, we will restrict our application of the model to the blebbistatin treated organoid with a less active core.

Clearly, two of the fit parameters in Figure~\ref{bio_systems}d-e are associated with an elliptical preferred shape ($e$ and $\phi$). That leaves five dimensionless parameters for a circular preferred shape: $\epsilon$, $c$, $k_t/\beta$, $T/a$, and $t_0/a$. (It is convenient to regard $T/a$, rather than $\sqrt{A_0}/a$, as a fitting parameter, as one may then guarantee Equation~\ref{lowerbound} is satisfied.) The first of these five is constrained by an experimental image from which one can measure (or at least estimate) the ratio of the $t$ and $r$ amplitudes. One can also measure $T$ and $\langle t\rangle$ in units of $\langle r\rangle$, as well as count the number of invaginations $q$. In general then, there are four independent constraints on these five model parameters. Suppose, however, the first measurement yields $\epsilon\ll1$, such as appears to be the case at the onset of shape change (around embryonic day 16.5 in mice~\cite{joy17}). In this limit, neither the leading order AC terms nor leading order DC offsets involve $c$, because it only appears in Equations~\ref{tsol}-\ref{rsol} as a product with $\epsilon$. Absent $c$, the remaining model parameters are all directly measureable: $k_t/\beta\approx q^2$, $T/a\approx T/\langle r\rangle$, $t_0/a\approx\langle t\rangle/\langle r\rangle$. 

The scaling behavior $q\sim\sqrt{k_t/\beta}$ can be understood in a simple way, as follows. As $\epsilon\to0$, the mechanical constraints on the outer surface of the fluid layer become relatively severe compared to those on the inner surface, because either the system-spanning radial springs are being turned into rigid rods ($k_r\to\infty$), or the core area constraint is being removed $(\mu\to0)$. Thus, the degrees of freedom representing different configurations of the outer surface are effectively frozen out of the inner surface problem. Only two terms now contribute significantly to the energy: the growth term, which scales as $E_{\textrm{grow}}\sim-k_tT^2$, and the gradient term, which scales as $E_{\textrm{grad}}\sim\beta T^2r^2/\lambda^2$. Here $\lambda$ is the wavelength of film thickness oscillations. Minimizing $E_{\textrm{grow}}+E_{\textrm{grad}}$ with respect to $T$ yields $\lambda\sim r\sqrt{\beta/k_t}$. Again, this scaling of the wavelength with system size is in contrast to elastic bilayer wrinkling, where the wavelength scales with film thickness. Higher-order growth instabilities lead to different scaling behavior just as different substrate models generate different scaling behavior in elastic bilayer wrinkling~\cite{cer03}. A lower order, hence linear, growth instability leads to unphysical results in the small $\epsilon$ limit, which is why we did not implement it here.

%%%
\section{Concave variant}
%%%

Suppose, instead of the convex bilayer depicted in Figure~\ref{schematic}, the system of interest is a concave bilayer contained within a rigid circular boundary having radius $R$. The substrate with conserved area $A_0$ is in contact with the boundary wall, and the growing (or shrinking) film is interior to that. Both layers are annular in shape, with thicknesses $r-t$ and $t$, respectively. Such a geometry has been used in prior work on morphogenesis of the gut~\cite{han11, shy13}, airways~\cite{wig97, li11}, and other tubular structures~\cite{zha17}. The sole difference in our formulation of this variant is that the variational principle involves an extra term with respect to the convex problem: $\delta\big(E-\mu\int d\theta [(r-t)^2 - 2R(r-t)] \big)=0$. One finds, in this case, that $R$ appears only in the DC offset terms of the solution
\begin{eqnarray}
t(\theta) &=& T\sin(q\theta + \phi) + \frac{(1-\epsilon)t_0+\epsilon c(a-R)}{1-\epsilon+\epsilon c} \nonumber\\
&+& \frac{\epsilon}{1-\epsilon}\bigg(\frac{k_rae^2}{2\beta}\bigg)\bigg(\frac{2-q^2\sin^2\theta}{(2-q^2)^2-4}\bigg),\label{tsol_concave}
\end{eqnarray}
and
\begin{equation}
r(\theta) = \bigg(\frac{1}{1-\epsilon}\bigg)r_0(\theta) - \bigg(\frac{\epsilon}{1-\epsilon}\bigg)[R+t(\theta)].\label{rsol_concave}
\end{equation}
Thus the phase diagram is the same as before, save for which regions correspond to growth and shrinkage. This latter difference is because
\begin{equation}
\frac{\langle t-t_0\rangle}{\langle r-t\rangle} = \epsilon c\bigg( \frac{\langle r_0-t_0\rangle - R}{\langle r_0-t_0\rangle + \epsilon(c-1)R} \bigg),
\end{equation}
which reduces to Equation~\ref{grow_shrink} only on the phase boundary $c=1-\epsilon^{-1}$. Note also that as $R\to\infty$, growth and shrinkage correspond to $c<1$ and $c>1$, respectively. The thickness amplitude is given by
\begin{equation}
T = \sqrt{2}(1-\epsilon)\sqrt{-\tfrac{A_0}{\pi} + \rho(2R-\rho) - \tfrac{a(R-\rho)}{1-\epsilon+\epsilon c}\tfrac{e^2}{2} + \mathcal{O}(e^4)},
\end{equation}
where $\rho = (a-t_0+\epsilon(c-1)R)/(1-\epsilon+\epsilon c)$, which implies an \emph{upper} bound on $A_0$.

In Figure~\ref{bio_systems}f, we plot a solution of this concave variant next to the image of the retinal fovea, as its (presumably differentially growing) layered structure is enclosed by a non-growing rigid shell on one side and vitreous gel on the other, while fibrous cells effectively span its ``cortex''. The parameters used (see the Figure caption) are consistent with $\langle t-t_0\rangle>0$, i.e., growth of the ganglion cell layer, despite the fact that $\epsilon$ is negative, which is a qualitative difference from the convex model. 

One may point out that our modeling of foveal pit morphogenesis is unrealistic because many pits are generated instead of one or two. But just as localized growth leads to localized buckling in elastic bilayer models, we can potentially resolve the discrepancy at hand by introducing a spatially inhomogeneous growth potential $k_t=k_t(\theta)$. Two analytic cases are deserving of mention. For the first case, suppose $k_t(\theta)$ is a piecewise periodic function, having a constant value in an arbitrarily small interval $[\theta_1,\theta_2]$ and a different constant value everywhere else. In the variational problem, we are free to take the bounds of integration as $\theta_1$ and $\theta_2$ (because there are no long-range interactions in the circumferential direction) and solve two independent problems: one for each growth potential region. Both solutions are of course given by Equations \ref{tsol_concave} and \ref{rsol_concave}, and it remains to apply matching boundary conditions. Unfortunately, taking $k_t\to0$ regionally leads to unphysical behavior in the analytic limits of the DC offsets (the same is true of the convex variant), so one must settle for a small but nonzero value of $k_t$. Alternatively, one might be interested in a slowly-varying growth potential, and in this case, we expect a local density approximation (LDA) to be valid. For small $\epsilon$, this would read $q(\theta)\approx\sqrt{k_t(\theta)/\beta}$. These possibilities for localized buckling also apply to the convex variant.

Finally, we suggest that it would also be interesting to apply this concave variant to the gut, for example, if the out-of-phase motif were discovered at the appropriate time of development.

%%%
\section{Normal thickness variant}
%%%

When there are strong deviations from circularity, the ``radial thickness" $t$ is no longer a good measure of film thickness. A more natural measure is the ``normal thickness" $\tau=t(\hat{r}\cdot\hat{n})$, where $\hat{r}=\cos(\theta)\hat{x} + \sin(\theta)\hat{y}$ is the unit radial vector and $\hat{n}=-\sin(\phi)\hat{x} + \cos(\phi)\hat{y}$ is the unit surface normal vector, with 
\begin{equation}
\phi(\theta,r,r') = \tan^{-1}\Big(\frac{r\cos\theta + r'\sin\theta}{r'\cos\theta - r\sin\theta}\Big).
\end{equation}
We now modify the variational problem (Equations~\ref{E} and \ref{varPrinc}), by replacing $t$ with $\tau$ in the last two terms of the energy functional
\begin{equation}
E^*[r,\tau,\tau'] = \int d\theta \Big{\{} k_r(r-r_0)^2 - k_t(\tau-t_0)^2 + \beta\Big(\frac{d\tau}{d\theta}\Big)^2 \Big{\}},\label{Enew}
\end{equation}
but leaving the variational principle otherwise unaltered:
\begin{equation}
\delta\Big(E^*-\mu\int d\theta\Big(r-\frac{\tau}{\hat{r}\cdot\hat{n}}\Big)^2\Big)=0.\label{varPrinc2}
\end{equation}
Choosing $r$ and $\tau$ as the independent variables, as we have done here, reduces the order of the Euler-Lagrange (EL) equations. Had we chosen instead to keep $r$ and $t$ as the independent variables, a higher derivative $r''$ would be introduced into the energy functional by the $(d\tau/d\theta)^2$ term, and fourth-order EL equations would result (see, for example, Ref.~\cite{cou53}). The reduced-order EL equations are given by
\begin{widetext}
\begin{eqnarray}
\tau'' &=& -\frac{\epsilon k_r}{N\beta}\Big(\frac{\tau}{N}-r\Big) - \frac{k_t}{\beta}(\tau-t_0),\label{tau''}\\
r'' &=& \frac{(r')^2}{r} + \frac{r^2+(r')^2}{r} \bigg( \frac{\frac{N}{\tau}\big[r^2+(r')^2\big]\big[\frac{1}{\epsilon}(\frac{r_0}{r}-1) - \frac{1}{r}(\frac{\tau}{N}-r) \big]+ \frac{M}{N}(\frac{2\tau'}{N} - r' - \frac{\tau' r}{\tau} )}{\frac{\tau}{N}(1+\frac{3M^2}{N^2}) - r(1+\frac{2M^2}{N^2}) + \frac{2Mr'}{Nr}(\frac{\tau}{N}-r)} \bigg),\label{r''}
\end{eqnarray}
\end{widetext}
where $N$ and $M$ are shorthand for the components of $\hat{r}$ in the direction of the surface normal and surface tangent, respectively, and $M/N=\frac{\tan\phi + \cot\theta}{1-\tan\phi\cot\theta}$. After simultaneously solving Equations~\ref{tau''}-\ref{r''}, the radial thickness may be recovered per its definition, $t=\tau/N$.

\begin{figure}[h]
\centering
\includegraphics[width=0.48\textwidth]{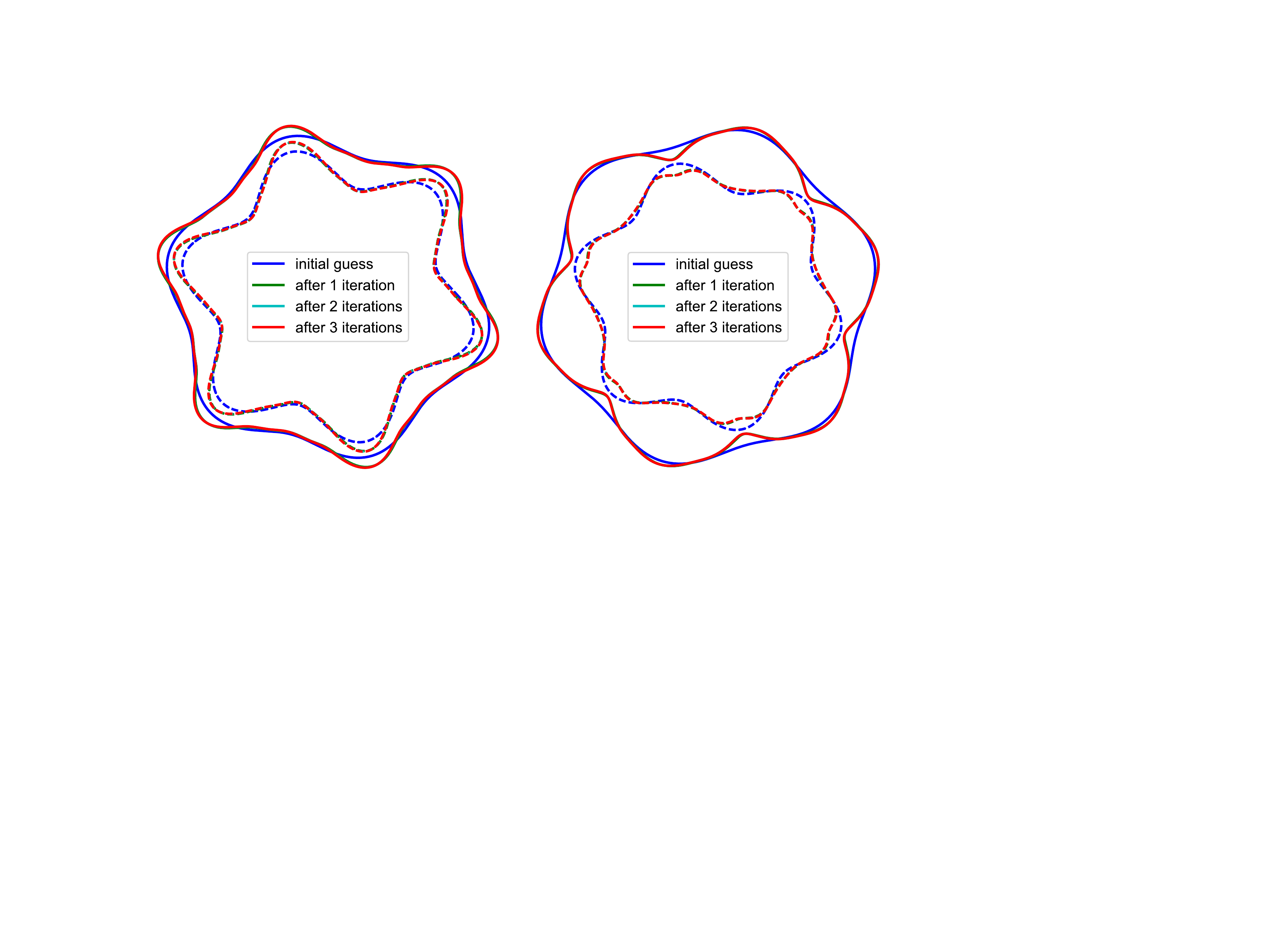}
\caption{\label{tau_variant}Numerical solutions of the nonlinear, normal thickness variant for selected cases of film growth (left panel) and film shrinkage (right panel). Dashed lines show $r-t$ and solid lines show $r$; colors indicate iterations. The zeroth iteration, shown in blue, is a solution of the linear model (Equations~\ref{tsol}-\ref{rsol}) with parameters as follows: $\epsilon=0.6$ (left panel) $-0.6$ (right panel), $c=T/a=0.1$, $t_0/a=0.2$, $q=6$, $\phi=e=0$.
}
\end{figure}

Though complicated and nonlinear, Equations \ref{tau''}-\ref{r''} are readily converted into a system of first-order difference equations, and solved numerically. We use multi-dimensional Newton-Raphson iteration on a periodic grid, with Krylov approximation of the Jacobian. The analytic solution of the corresponding linear problem provides a convenient initial guess for relaxation. While these numerical solutions show some sensitivity to the number of gridpoints used, we find that 10,000 gridpoints gives rapid convergence and reproduces the linear model's behavior at small $\epsilon$, i.e., where the radial thickness and normal thickness definitions coincide. Furthermore, at modest values of $\epsilon$ and $c$, the observed deviation of the shapes from those of the linear model can be rationalized as follows. A sharp bend in a film having uniform normal thickness incurs no penalty from a $(d\tau/d\theta)^2$ term, but it does incur a penalty from a $(dt/d\theta)^2$ term (although not in the same sense as that for bending an elastic film). Therefore, we might expect the normal thickness variant to exploit the extra degrees of freedom afforded by these low energy, sharp bends in negotiating the competition between growth/shrinkage and preferred radius. Figure~\ref{tau_variant} shows these sharp bends occur preferentially where the film is thin. One interesting consequence of this behavior is that additional minima in the substrate radius can be introduced, e.g., in the right panel of Figure~\ref{tau_variant}. In contrast, where the film is thick, its shape is not significantly distorted from that in the corresponding linear model.
%
% Figure 5
%
%
%
%

%%%
\section{Discussion}
%%%

We have demonstrated that the out-of-phase behavior observed in certain morphogenesis contexts is not only at odds with an elastic bilayer wrinkling mechanism, but indicative that microstructural details affect morphogenesis in heretofore unappreciated ways. Simply treating the problem as one of differential growth between two homogeneous elastic materials appears insufficient to capture the unique shapes of the developing cerebellum, +blebbistatin organoid, and the retinal fovea. We have constructed a minimal model that captures these shapes (at least qualitatively), and in our model radially oriented fibers play a key role. 

It is reassuring, then, to notice that all three of the above mentioned biological systems contain one or more types of radial fibers. One is therefore led to speculate whether our model is solely applicable to the central nervous system, where requisite fibers such as glia are present during development of all regions, and certain of whose organs exhibit the telltale out-of-phase behavior. At the time of writing, we are not aware of definitive out-of-phase behavior in other morphogenesis contexts, but we are actively searching. The developing \emph{cerebrum} is a natural place to look, but interestingly, in adult humans, the cerebral cortex thickness appears to be in-phase with its surface height (see Figure 1 in the Supplementary Information of Ref.~\cite{gla16}). Another place to look is where there is smooth muscle. If the smooth muscle tissue is very thin and minimally connected, presumably the fibrous nature of the cells making up the tissue call for a micro-scale mechanical description like the one we have developed here. A novel modeling description for such a system could lead to additional novel shape changing mechanisms for developing organs.

Application of the model to other developing organs may also potentially call for an extension into the third dimension.  One can then define a spherical inner core and a spherical outer shell of proliferating cells as well as fibrous cells radially extending themselves throughout the system.  The developing undulations along the perimeter of the shape in two-dimensions now becomes developing undulations over a surface and one can ask whether or not the undulations on the outer surface are in-phase or out-of-phase with the inner surface, depending on the parameters. Since the key new aspects of the model are all contained within the two-dimensional version, we focus on its results for now.

The possibility of constructing a synthetic device that embodies the Hamiltonian given by Equations~\ref{E}-\ref{constraint} is an intriguing one. Such a device would consist of (1) an incompressible core, (2) a growing film that has the curious combination of flexibility in bending and stiffness against thickness gradients, and (3) system-spanning, radially oriented, elastic fibers. (See the discussion in Ref.~\cite{law18} for an alternative, potentially simpler way to realize the $\sim k_r$ term.) The film component would likely have to be some sort of fiber-matrix composite material, as it also appears to be in the cerebellum. The nontrivial mechanical properties of this film component could perhaps be reverse engineered by trying different composite formulations and watching for out-of-phase behavior.

Potential applications of the model outside of molecular and tissue-scale biology may also be worth investigating. Morphogenesis of large cities, for example, shares several of the key ingredients of our model (at least in a qualitative sense). Cities are 2d structures, and there is in certain cases an ``incompressible" urban core, surrounded by a growing suburban belt. The preferred radius concept is also not unreasonable: people relocating to or within the city might be expected to strike a balance between commuting time to the city center and housing prices, which tend to fall off with distance from the center. Chengdu, the capital of China's Sichuan province, is one example of a large and rapidly growing city that is free of significant geological constraints on its shape changes (as evidenced by its nearly circular shape). It would be interesting to try and interpret Chengdu's shape changes within the past few decades~\cite{sch15} from within a differential growth framework.

Model variants beyond the two described here are of course possible, as well as numerous. Already mentioned was the notion of putting spatial-dependence into one or more parameters. One could also, for example, introduce a curvature-dependent growth potential, $k_t=k_t(r'')$, which might be expected to control the degree of cuspiness of the invaginations. An anharmonic correction to the preferred radius term, or a tight-packing constraint, might generate self-contacting folds by squashing or squeezing the lobules (i.e., wrinkle crests) together. Neither the linear model nor the normal thickness variant that we have investigated here appear capable of producing cusped invaginations or self-contacting folds, such as those that occur in mouse cerebellum development after about 18 embryonic days~\cite{joy17}. Extending the model beyond just the onset of shape changes, to the large growth and deformation regimes, will open up the possibility of new kinds of tests, including  whether or not it can capture subdivision of lobules. 

The preferred radius $r_0$ naturally suggests a mechanism for subdivisions, and as previously mentioned, subdivisions occur in the mammalian cerebellum during later development. Time-dependence could be introduced into our quasi-static model or one of its variants in such a way that lobules are dynamically growing in area. When a lobule's area reaches a value $\sim r_0^2$, the lobule may in fact constitute a subsystem that resembles the full system at an earlier point in time. Shape change of the subsystem would be akin to a folding hierarchy, and one might imagine this going on for several generations. An implementation in which subdivisions emerge spontaneously, as opposed to being put in ``by hand", is a very interesting direction for future work. 

Finally, we point out that an apparent scale-invariance in the wavenumber $q$, found here in the limit $\epsilon\to0$, is reminiscent of ``Larsell's criterion for the vermis"~\cite{sil07}. Adult mammalian cerebella span two orders of magnitude in size, ranging from $\sim1$ mm in mouse to $\sim10$ cm in humans, and Larsell observed/argued that these can all be considered as having an underlying 10-fold motif. In his book, Larsell notes that other researchers before him including Bradley, Bolk, and Riley conducted comparative anatomy studies of the cerebellum (an embryological study spanning six species in the case of Bradley) and drew similar conclusions about its scale-invariance~\cite{lar70}. While Larsell's criterion is not universally accepted by biologists, the fact remains that the cerebellum of all mammal species is highly folded, whereas the cerebrum is unfolded (i.e., lissencephalic) in the smallest mammals, thus there is at least a hint of scale-invariance in cerebellar morphogenesis, which our model may capture.

%%%
\section{biological perspective}
%%%

A biologist might naturally ask: if buckling is a mechanical phenomenon that requires no genetic prepatterning, then why does the specific buckling mechanism (\emph{with} versus \emph{without} bending) matter for biological contexts? What can I infer about the biology of a tissue from the phase and amplitude behavior of its wrinkles?

In our view, these kinds of questions (many thanks to an anonymous referee for posing them), target the notion of emergent behavior in bulk material systems. To use a classic example, some bulk materials superconduct at low temperatures while others don't -- these differences in emergent behavior can teach us something about the basic atomic building blocks of the materials and the way those blocks are arranged, via, for example, the electron-phonon interaction. This is ``backwards" from the reductionist viewpoint, that would have us start from the basic atomic building blocks to try and understand superconductivity.

Likewise, in organ morphogenesis, the macroscopic phase, amplitude, and wavelength behaviors are associated with the emergent phenomenon of buckling instability, and they can teach us something about the micro-biological building blocks. In-phase behavior (and other behavior consistent with the rightmost column of Table~\ref{summaryFigure}) teaches us the building blocks are effectively homogeneous, elastic materials. In other words, if there are multiple cell types present, they should be mixed together on a fine length-scale in every direction. Furthermore, there must be little to no stress-relieving rearrangements and neighbor exchanges of cells (i.e., fluid-like behavior) on the $\sim12$ hour timescale of shape changes; all significant stress reduction happens over the much larger length-scale of the wrinkles. Out-of-phase behavior teaches us that certain types of fiber-like cells are present and span relatively long length scales, and further suggests what will happen when we add or remove these fibers genetically (see the consequences of tuning $k_r$ and $\beta$ summarized in Table~\ref{summaryFigure}). It also provides some clues as to what is happening to the fibers as the organ grows over time (see our experimental companion paper, Ref.~\cite{law18}, which further tests both models against mouse cerebellum development, including predictions beyond those listed in Table~\ref{summaryFigure}). Finally and perhaps most importantly, out-of-phase behavior is indicative of fluid-like rearrangements and neighbor exchanges of at least one cell type in the cortex/film component, presumably as a result of both cell divisions and cell motility. Such lessons from emergence may ultimately help us understand and correct what is going wrong, micro-biologically, in brain folding-related diseases such as lissencephaly and polymicrogyria, and in certain malformations of the retina such as the bifoveality shown in Figure~\ref{bio_systems}c.
%
% Table 1
%
\begin{table*}[t]
\centering
\caption{Summary of main predictions for the growth regime ($0<\epsilon<1$) of the convex variant, with comparison to the corresponding predictions of a standard morphogenesis model. The $\epsilon\ll1$ limit pertains to early development, i.e., just after the onset of shape change. The five dimensionless model parameters discussed above are written in a slightly different (but equivalent) form here in order to emphasize their physical meaning.\label{summaryFigure}}
{\setlength{\tabcolsep}{0.2cm} 
\renewcommand{\arraystretch}{2.0}
\begin{tabular}{|p{6.0cm}|p{5.35cm}|p{5.35cm}|}
\hline
%%% row 1
& \parbox[c]{5.35cm}{\bf{BUCKLING WITHOUT BENDING MODEL}} & \parbox[c]{5.35cm}{\bf{CONVENTIONAL ELASTIC BILAYER MODEL}} \\[0.2cm]
\hline
%%% row 2
\parbox[c]{6.0cm}{phase relationship between film thickness and substrate deformation (the planar limit $t/r\to0$ is shown for simplicity)} & \parbox[c]{5.35cm}{\vspace{1pt}\includegraphics[scale=0.4]{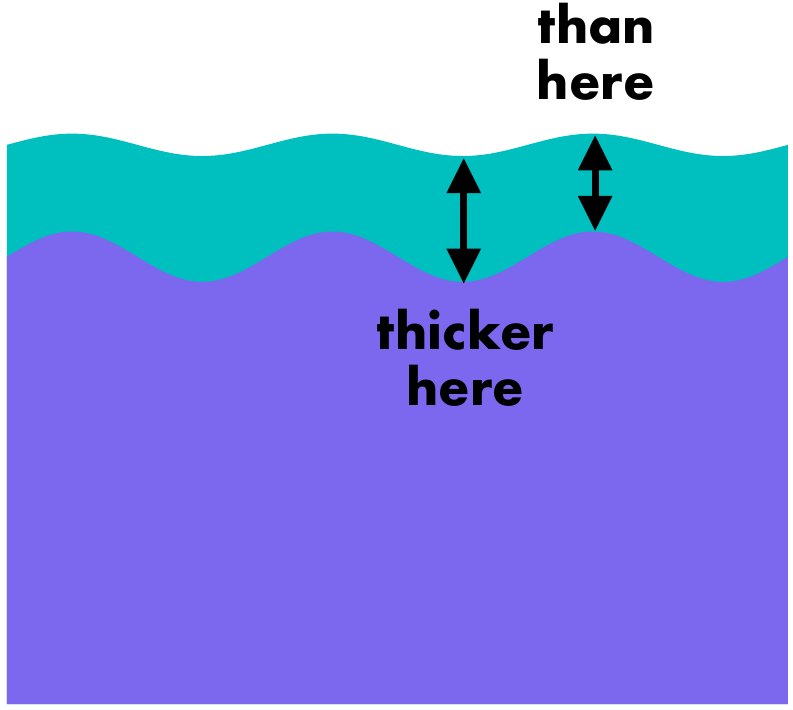}} & \parbox[c]{5.35cm}{\vspace{1pt}\includegraphics[scale=0.4]{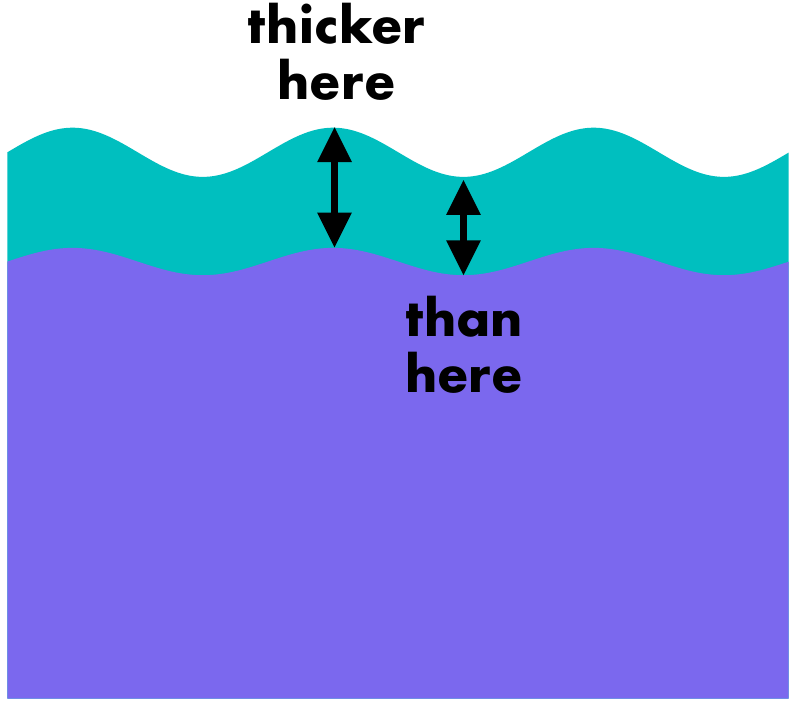}} \\[1.25cm]
\hline
%%% row 3
\parbox[c]{6.0cm}{amplitude of film thickness oscillations} & \parbox[c]{5.35cm}{$=\frac{1-\epsilon}{\epsilon}\,\times$ wrinkling amplitude} & \parbox[c]{5.35cm}{$\ll$ wrinkling amplitude} \\[0.2cm]
\hline
%%% row 4
\parbox[c]{6.0cm}{wavenumber $q=\frac{2\pi\langle r\rangle}{\lambda}$} & \parbox[c]{5.35cm}{$=\sqrt{\frac{k_t}{\beta}\big(1+\frac{\epsilon c}{1-\epsilon}\big)} \approx \sqrt{\frac{k_t}{\beta}}$, for $\epsilon\ll1$} & \parbox[c]{5.35cm}{$\sim\frac{\langle r\rangle}{t}\Big(\frac{E_f}{E_s}\Big)^{-1/3}$} \\[0.2cm]
\hline
%%% row 5
\parbox[c]{6.0cm}{near morphogenesis onset, the number of wrinkles $q$ is independent of} & \parbox[c]{5.35cm}{geometry} & \parbox[c]{5.35cm}{differential strain (in excess of critical strain)} \\[0.2cm]
\hline
%%% row 6
\parbox[c]{6.0cm}{to generate more wrinkles} & \parbox[c]{5.35cm}{increase growth potential $k_t$ or decrease gradient penalty $\beta$} & \parbox[c]{5.35cm}{decrease film thickness (relative to system size) or decrease stiffness contrast $E_f/E_s$} \\[0.2cm]
\hline
%%% row 7
\parbox[c]{6.0cm}{proxy for time in developmental dynamics} & \parbox[c]{5.35cm}{$\epsilon$ (see Ref.~\cite{law18} for more details)} & \parbox[c]{5.35cm}{differential strain} \\[0.2cm]
\hline
%%% row 8
\parbox[c]{6.0cm}{minimal input physics in the form of dimensionless parameters} & \parbox[c]{5.35cm}{
\begin{itemize}[leftmargin=*]
\item effective radial spring constant $k_r/\mu$ $(=\epsilon^{-1})$ presumably coming from system-spanning radial glia fibers 
\item growth potential $k_t/\mu$ of the film
\item thickness gradient penalty $\beta/\mu$ presumably coming from film-spanning fibers, e.g., Bergmann glia
\item preferred geometry $\sqrt{A_0}/r_0$
\item reference geometry $t_0/r_0$
\end{itemize}}
& \parbox[c]{5.35cm}{
\begin{itemize}[leftmargin=*] 
\item stiffness contrast $E_f/E_s$ of two homogeneous elastic materials 
\item zero-strain geometry $\tau^0/r_s^0$ 
\item differential strain $e_{\theta\theta}$
\end{itemize}
}\\[0.2cm]
\hline
\end{tabular}}
\end{table*}
%
%
%

% Specify following sections are appendices. Use \appendix* if there
% only one appendix.
\appendix*
\section{Finite Element Simulations}
We performed finite element (FE) simulations for wrinkling in planar and circular bilayered
structures (i.e., a thin film bonded on a substrate) to investigate the thickness variation in the
buckled film. The bilayer structures were assumed to be under 2d plane strain deformation. The
elastic properties of both film and substrate are described by the incompressible neo-Hookean
model, whose strain energy can be expressed as
\begin{equation}
U = \frac{1}{2}G(I_1-3),
\end{equation}
where $G$ is the shear modulus and $I_1$ represents the first invariant of the right Cauchy-Green deformation tensor. All of the FE simulations were performed with ABAQUS standard solver. The
bilayered structures were discretized with CPE6MH element with the smallest mesh size less
than 10\% of the film thickness. To simulate the buckling behaviors due to differential growth in
the film and substrate, an isotropic growth deformation was applied to the film through the
``expansion" function in ABAQUS. A random perturbation (white Gaussian noise with mean
magnitude equal to 0.1\% of the initial film thickness) was applied to the nodal positions at the
top surface to trigger the wrinkle instability.

\begin{figure}[b]
\centering
\includegraphics[width=0.46\textwidth]{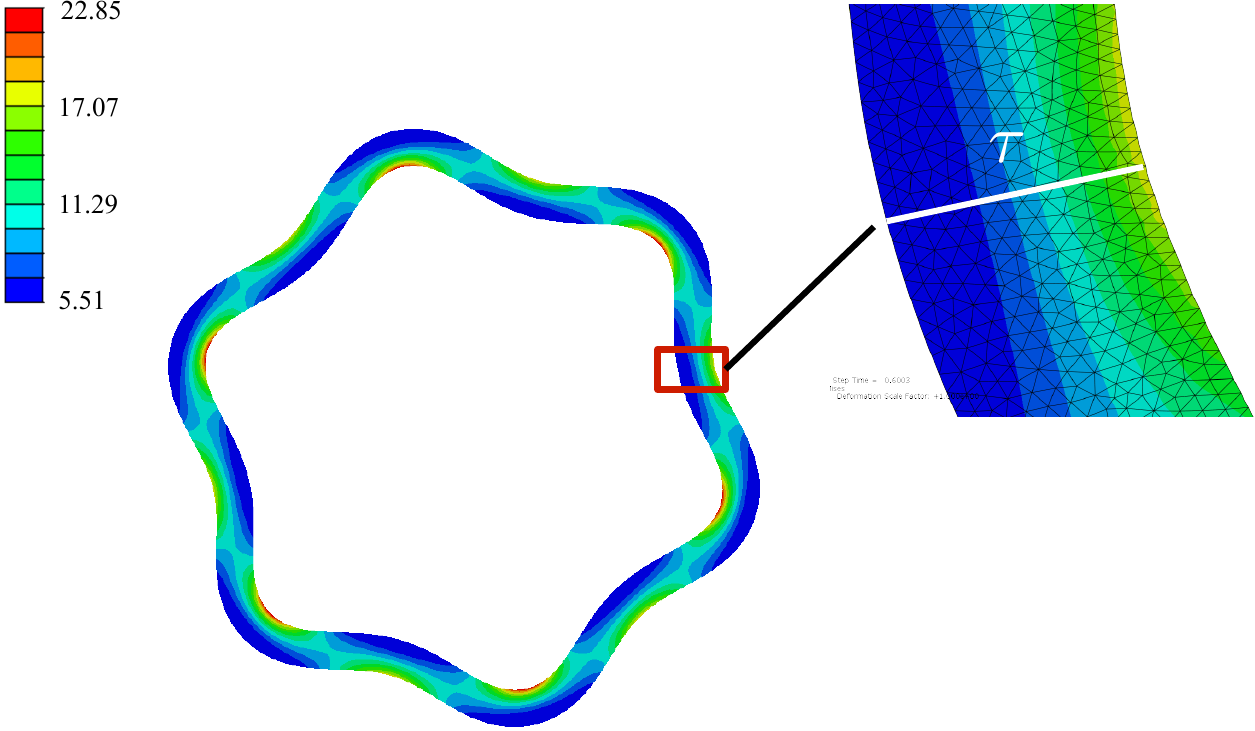}
\caption{\label{SI_f1}Detail of normal thickness $(\tau)$ measurement. Color represents the von-Mises stress.}
\end{figure}

\begin{figure*}[t]
\centering
\includegraphics[width=0.75\textwidth]{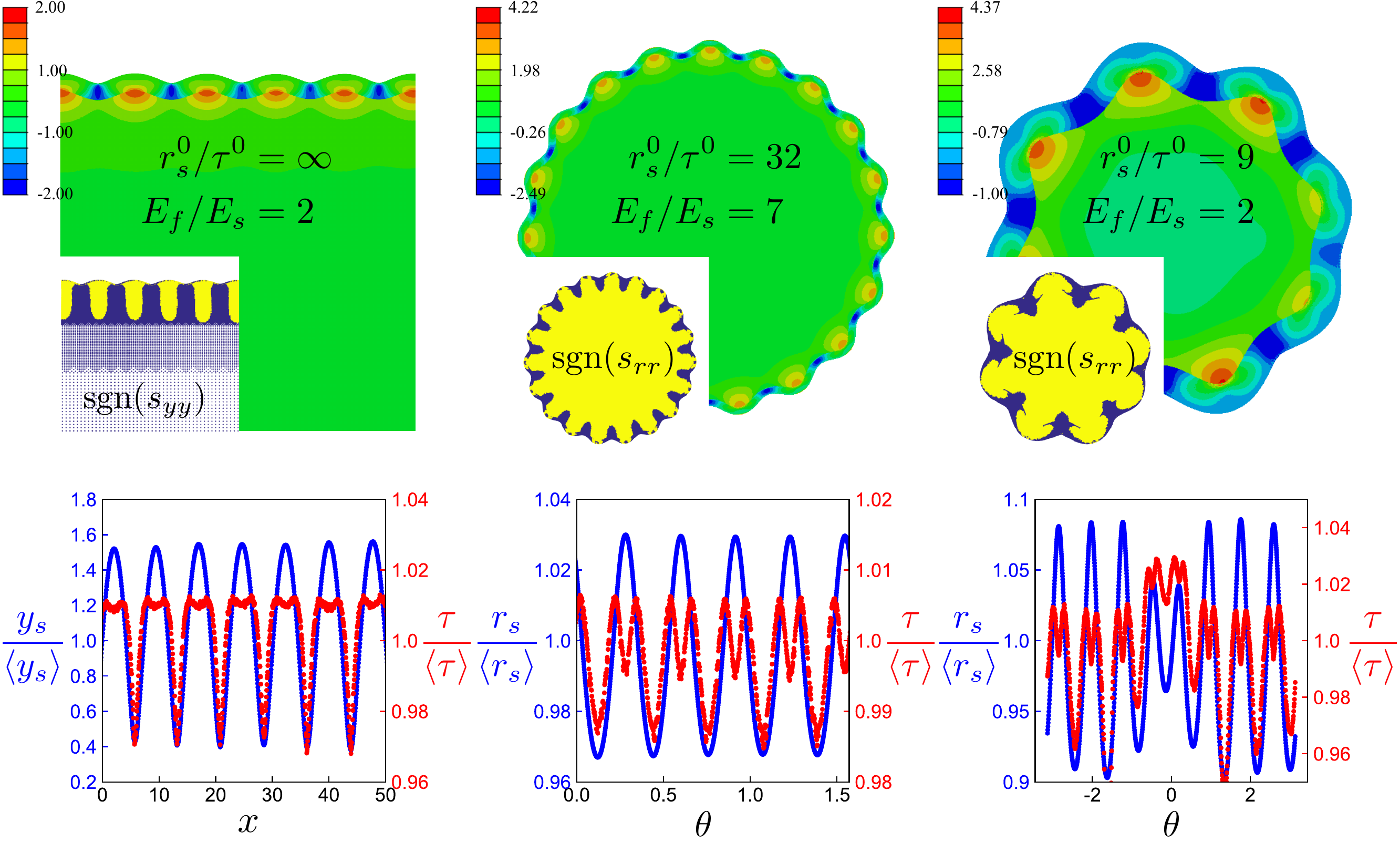}
\caption{\label{SI_f2}Additional bilayer wrinkling simulations, similar to those shown in Figure~\ref{circular_wrinkle}, but with different geometries and modulus ratios. Top row: maximum in-plane principal stress normalized to half of the substrate's shear modulus. Insets show the sign of the stress tensor component in the direction of the surface normal, with yellow indicating positive values (tension) and purple indicating negative values (compression). Bottom row: substrate height (blue) and film thickness (red), normalized to their average values. The rightmost case with a relatively thick film shows a rotational symmetry breaking that may be indicative of global buckling.}
\end{figure*}

The film thickness in the wrinkled state is approximated as the shortest distance between the FE
nodes at the inner and outer surface of the film, as shown in Figure~\ref{SI_f1}. We observe some noise in
these thickness data, which can be attributed to the finite size of the FE mesh. However, this
noise is a second order perturbation and will not change the overall thickness variation of the
film. The film thickness variation is compared to the substrate deformation in Figure~\ref{SI_f2}.
%
% Figure 6
%
%
% Figure 7
%

\clearpage

% If you have acknowledgments, this puts in the proper section head.
\begin{acknowledgments}
TAE and JMS thank Ellen Kuhl, Johannes Weickenmeier, Orly Reiner, and Helga Kolb for useful discussions, and acknowledge financial support from NSF-DMR-CMMT Award Number 1507938 and NSF-PHY-PoLS Award Number 1607416. FE simulations were performed at the Comet cluster (Award no. TGMSS170004) in XSEDE. Additional financial support is acknowledged from NIMH-R37MH085726 and NINDS-R01NS092096 to ALJ and F32 NS086163 to AKL, and a National Cancer Institute Cancer Center Support Grant [P30 CA008748-48].
\end{acknowledgments}

% Create the reference section using BibTeX:
%\bibliography{CBmodel_withTitles}
\input{morphogenesis13.bbl}
\end{document}

%% file: morphogenesis13.bbl
%merlin.mbs apsrev4-1.bst 2010-07-25 4.21a (PWD, AO, DPC) hacked
%Control: key (0)
%Control: author (0) dotless jnrlst
%Control: editor formatted (1) identically to author
%Control: production of article title (0) allowed
%Control: page (1) range
%Control: year (0) verbatim
%Control: production of eprint (0) enabled
\providecommand{\noopsort}[1]{}\providecommand{\singleletter}[1]{#1}%